# Superradiance of non-Dicke states

N. E. Nefedkin,[1,2,3] E. S. Andrianov,[1,2] A. A. Zyablovsky,[1,2] A. A. Pukhov,[1,2,3] A. P. Vinogradov,[1,2,3] and A. A. Lisyansky[4,5]

[1] *Dukhov Research Institute for Automatics, 22 Sushchevskaya, Moscow 127055, Russia*
[2] *Moscow Institute of Physics and Technology, 9 Institutskiy per., Dolgoprudniy 141700, Moscow Reg., Russia*
[3] *Institute for Theoretical and Applied Electromagnetics RAS, 13 Izhorskaya, Moscow 125412, Russia*
[4] *Department of Physics, Queens College of the City University of New York, Queens, NY 11367, USA*
[5] *The Graduate Center of the City University of New York, New York, New York 10016, USA*

In 1954, Dicke predicted that a system of quantum emitters confined to a subwavelength volume would produce a superradiant burst. For such a burst to occur, the emitters must be in the special Dicke state with zero dipole moment. We show that a superradiant burst may also arise for non-Dicke initial states with nonzero dipole moment. Both for Dicke and non-Dicke initial states, superradiance arises due to a decrease in the dispersion of the quantum phase of the emitter state. For non-Dicke states, the quantum phase is related to the phase of long-period envelopes which modulate the oscillations of the dipole moments. A decrease in dispersion of the quantum phase causes a decrease in the dispersion of envelope phases that results in constructive interference of the envelopes and the superradiant burst.

## 1. INTRODUCTION

Superradiance (SR) is a sharp enhancement of the spontaneous radiation rate of an ensemble of $N$ independent emitters (two-level atoms) compared to the radiation rate of a single emitter, $\gamma_0$. This phenomenon was predicted by Dicke [1] for a subwavelength collection of $N$ quantum emitters that are coupled by their own radiation field. Various aspects of this phenomenon are reviewed in Refs. [2-9].

Dicke assumed that all emitters are indistinguishable and their wave function is symmetric with respect to permutations of any two emitters. In a general form, the Dicke state of $N$ two-level atoms, $n$ of which are excited, has the form

$$|\psi_D\rangle \equiv |N,n\rangle = \frac{1}{\sqrt{C_N^n}} \sum_P \left| \underbrace{e,...,e}_{n}, \underbrace{g,...,g,}_{N-n} \right\rangle, \qquad (1)$$

where *P* denotes all possible permutations. As an initial state, Dicke considered a state in which all *N* emitters are excited [3,4]. The dipole moment of such a system is equal to zero. Dicke took into account only one channel of the system's evolution in which at each time step only one of the emitters relaxes to the ground state and the system proceeds to another pure Dicke state $|N,n-1\rangle$. Thus, at any time, the dipole moment remains equal to zero. In the Dicke model, the probability of the transition from the state with *n* to *n* –1 excited emitters per unit time (the radiation rate), $\gamma(n)$, depends on *n*. In the initial moment, when all emitters are excited, $\gamma(n)$ is at a minimum, $\gamma(N) = \gamma_0 N$ [1,2,4]. It reaches a maximum value of $\gamma_0 N^2/4$ when $n = N/2$. Thus, when half of the emitters are excited, the radiation rate depends quadratically on the number of emitters, while initially, this dependence is linear. This increase in the radiation rate of two-level atoms is characteristic to SR. Dicke showed that the peak in the radiation intensity is reached in a time $\sim \log N/N$, while the duration of the SR burst is smaller than the radiation time of a single emitter by a factor of $1/N$.

SR may arise for any Dicke state with $n \leq N$ excited atoms. For example, an SR state can be a state with a single excited atom, which is symmetric with respect to all possible permutations [10-12]. However, SR depends strongly on the initial state. When the initial state is antisymmetric with respect to atomic permutations, instead of SR, radiation is suppressed. It becomes even smaller than radiation of independent emitters. This phenomenon is called *subradiance* [4].

The Dicke explanation of SR is based on a strong assumption about the time evolution from one Dicke state to another (only in this case can one use the Fermi's golden rule) and on an ability of a quantum system with zero dipole moment to radiate a photon. There is no rigorous proof that these conditions are either necessary or sufficient for SR. Moreover, it has been shown that both of these conditions are not quite correct.

First, a more rigorous description of SR in terms of the master equation shows that the time evolution does not go through pure Dicke states but rather through mixed states with a density matrix which is a linear combination of density matrices of various pure Dicke states $\rho_{mixed} = \sum_n c_n |\psi_D\rangle\langle\psi_D|$ [4]. These mixed states still have a zero dipole moment. Second, SR is not the sole prerogative of quantum systems. It can also occur in an ensemble of nonlinear classical oscillators, which surely has a nonzero dipole moment [13,14]. In such a system, SR results from the constructive interference of long-period envelopes of rapidly oscillating dipoles [13]. For this reason, it is interesting to investigate whether SR may occur from quantum states with nonzero dipole moments, which are not Dicke states. It is worth noting that there are several phenomena discussed in the literature which one way or another are similar to SR. These are superfluorescence [15,16], superluminescence [17,18], and amplified spontaneous emission (ASE) [19-21]. Some of the conditions required for the observation of these phenomena are the same as for SR. We want to emphasize that we consider a subwavelength system of quantum emitters that are initially excited non-coherently. The focus of our study is the effect of the initial dipole moment of the system dynamics.

In this paper, we study the possibility of SR in an ensemble of two-level atoms in the general case, in which the system is not initially in a Dicke state. We show that for a quantum system, there is a unified mechanism for SR for both Dicke states with zero dipole moment and non-Dicke states for which the total dipole moment is not zero. We introduce a phase operator for a quantum state and show that this mechanism is related to a decrease in the dispersion of the state phase. The SR burst occurs when the dispersion reaches its minimum value. The expectation value of the initial dipole moment only affects the time delay. The greater the expected value, the smaller the time delay. We also show that nonlinearity is essential for SR to arise.

## 2. THE DICKE MODEL OF SUPPERRADIANCE

Let us first briefly consider the Dicke theory (see for details Refs. [3,4]). For an ensemble of $N$ two-level atoms we introduce the lowering and raising operators that describe the relaxation and excitation of the $j$-th atom $\hat{\sigma}_j = \underbrace{\hat{E} \otimes \hat{E} \otimes ... \otimes}_{j-1} \hat{\sigma} \underbrace{\otimes ... \otimes \hat{E}}_{N-j}$ and $\hat{\sigma}_j^+ = \underbrace{\hat{E} \otimes \hat{E} \otimes ... \otimes}_{j-1} \hat{\sigma}^+ \underbrace{\otimes ... \otimes \hat{E}}_{N-j}$, where $\hat{\sigma} = |g\rangle\langle e|$ and $\hat{\sigma}^+ = |e\rangle\langle g|$ are transition operators from excited $|e\rangle$ and ground $|g\rangle$ states, respectively, and $\hat{E}$ is the $2\times 2$ identity matrix. The corresponding Hamiltonian of the Jaynes-Cummings type for the interaction between free-space field modes and two-level atoms in the rotating wave approximation is:

$$\hat{H} = \sum_k \hbar\omega_k \hat{a}_k^+ \hat{a}_k + \hbar\omega \hat{J}^z + \sum_k \hbar\Omega_k \left( \hat{a}_k^+ \hat{J}^- + \hat{J}^+ \hat{a}_k \right) \quad (2)$$

where $\hat{a}_k^+$ and $\hat{a}_k$ are creation and annihilation operators of a photon in a mode with the frequency $\omega_k$, $\omega$ is the transition frequency of two-level atoms, $\Omega_k$ is the interaction constant between photons and atoms, $\hat{J}^- = \sum_j \hat{\sigma}_j$ is the collective atomic operator of the complex dipole moment, $\hat{J}^+ = \left(\hat{J}^-\right)^+ = \sum_j \hat{\sigma}_j^+$, and $\hat{J}^z = \sum_j \hat{\sigma}_j^z$ stands for the collective inversion, where $\hat{\sigma}_j^z = \underbrace{\hat{E} \otimes \hat{E} \otimes ... \otimes}_{j-1} \hat{\sigma}^z \underbrace{\otimes ... \otimes \hat{E}}_{N-j}$ and $\sigma^z = |e\rangle\langle e| - |g\rangle\langle g|$ [4].

Using the Heisenberg approach and the integral of motion, $\left(\hat{J}^+\hat{J}^- + \hat{J}^-\hat{J}^+\right)/2 + \left(\hat{J}^z\right)^2/4$, one can eliminate the variable $\hat{J}^-$. The remaining equation for $\hat{J}^z$ is an operator equation which should be converted into an equation for the expectation value $\langle \hat{J}^z \rangle$, where $\langle ... \rangle$ denotes an average value of an operator calculated as $\langle \hat{J}^z \rangle = \text{Tr}\left(\rho(t)\hat{J}^z\right)$, where $\rho(t)$ is the density matrix. The Markovian approximation allows for the elimination of the field variables $\hat{a}_k^+$ and $\hat{a}_k$ [3,4]. At this stage, the rate $\gamma_0$ of the spontaneous radiation into free-space modes is introduced. This parameter sets the characteristic time-scale $\gamma_0^{-1}$. After exclusion of the field variables, only one variable, $\hat{J}^z$, remains. For the Dicke state $\langle \hat{J}^z \rangle = n - (N-n) = 2n - N$. At this point, the second approximation, $\langle \hat{J}^z \hat{J}^z \rangle = \langle \hat{J}^z \rangle \langle \hat{J}^z \rangle$, should be made. This is correct when $N \gg 1$. As a result, one obtains that the dynamics of the collective inversion $\hat{J}^z = \sum_i \hat{\sigma}_i^z$ of a system that is in a pure state (1) at any moment of time, can be described by the original Dicke equation

$$\frac{d\langle \hat{J}^z \rangle}{dt} = -\gamma_0 \left( N^2/4 + N/2 - \frac{1}{4}\langle \hat{J}^z \rangle^2 + \frac{1}{2}\langle \hat{J}^z \rangle \right). \quad (3)$$

Solving this equation Dicke obtained the time dependence of the inversion:

$$\langle \hat{J}^z(t) \rangle = 1 - (N+1)\tanh\left(\gamma_0(N+1)(t - t_{delay})/4\right), \quad (4)$$

where $t_{delay}$ is determined from the initial condition $\langle \hat{J}^z(0) \rangle = N$. It is equal to

$$t_{delay} = \frac{2\ln N}{\gamma_0(N+1)}. \quad (5)$$

The radiation intensity has a form of a burst

$$I(t) = -d\langle J^z(t)\rangle/dt = \gamma_0 \left(\frac{N+1}{2}\right)^2 \text{sech}^2\left(\gamma_0(N+1)(t-t_{delay})/4\right) \quad (6)$$

From Eq. (6) one can see that the intensity maximum occurs at $t = t_{delay}$. Thus, $t_{delay}$ has a meaning of the delay time of the SR burst. Note, that at the initial moment, the system inversion is $\langle \hat{J}^z(0)\rangle = N$, while in $t_{delay}$, according to Eq. (4), $\langle \hat{J}^z(t_{delay})\rangle = 1$, i.e., at this moment $n \approx N/2$. Thus, the SR burst arises when about half of the atoms are excited.

Expression (5) for the delay time for the SR burst follows from the solution of the Dicke equation (3). This time can be found by assuming that the system evolution is going through Dicke states. Indeed, this time is comprised of the step-by-step transition times from the state $|N,N\rangle$ to the state $|N,N/2\rangle$. According to the Fermi's golden rule, the probability of the transition from the state $|N,n\rangle$ to the state $|N,n-1\rangle$ per unit time (the radiation rate) is $\gamma(n) = \gamma_0 n(N-n+1)/2$ [1,3,4]. The average transition time between these states is $\gamma(n)^{-1}$. Then the average transition time from the initial state $|N,N\rangle$ to the state $|N,N/2\rangle$ can be estimated as

$$T_{|N,N\rangle \to |N,N/2\rangle} = \sum_{n=N}^{n=N/2} \gamma(n)^{-1} = \frac{2}{\gamma_0} \sum_{n=N}^{n=N/2} \frac{1}{n(N-n+1)} \approx \frac{2}{\gamma_0(N+1)} \sum_{n=N}^{n=1} \left(\frac{1}{n}\right) \approx \frac{2\ln N}{\gamma_0(N+1)}. \quad (7)$$

This expression is identical to expression (5) obtained by solving Dicke equation (3).

If the initial number of excited emitters $n$ in the Dicke state is smaller than $N$, then the transition time to the $|N,N/2\rangle$ state decreases because of the decrease in the number of terms in the sum in Eq. (7). As the number of excited emitters approaches $N/2$, the delay time tends to zero. This is confirmed by our computer simulations shown in Fig. 1 and is in agreement with the results of Refs. [10-12] in which the case $n = 1$ has been considered.

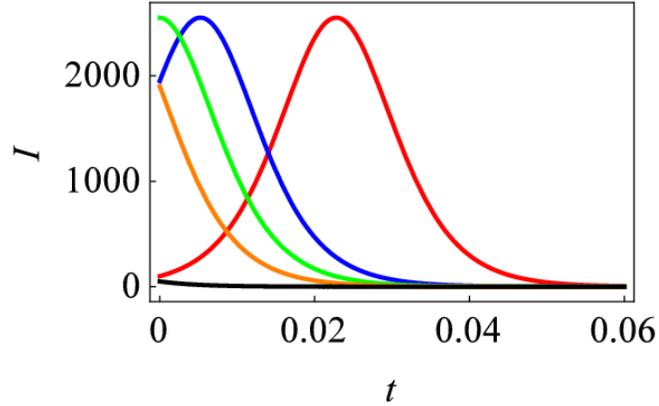

Fig. 1. The radiation intensity in the Dicke model as a function of time for a different number of initially excited atoms. $n = N$ (the red line), $n = 0.75N$ (the green line), $n = 0.5N$ (the blue line), $n = 0.25N$ (the orange line), and $n = 0.01N$ (the black line).

The Dicke system can also be described in a different way without using the Dicke assumptions. If we restrict ourselves to the case of $N\gamma_0 \ll \omega$, then the dynamics of the system can be described by the density matrix governed by the Lindblad master equation [4,22,23]:

$$\dot{\rho} = -\frac{i}{2}\omega_A \sum_{j=1}^{N}\left[\hat{\sigma}_j^z, \rho\right] + \frac{\gamma_0}{2}\sum_{i,j=1}^{N}\left(2\hat{\sigma}_i \rho \hat{\sigma}_j^+ - \hat{\sigma}_i^+ \hat{\sigma}_j \rho - \rho \hat{\sigma}_i^+ \hat{\sigma}_j\right), \tag{8}$$

where $[,]$ denotes a commutator of the respective operators. Using the interaction representation, $\rho \to \exp\left(i\omega t \sum_j \hat{\sigma}_j^z/2\right)\rho \exp\left(-i\omega t \sum_j \hat{\sigma}_j^z/2\right)$, we consider smooth oscillations (envelopes). Switching from single-particle operators, $\hat{\sigma}_j$ and $\hat{\sigma}_j^+$, to collective operators, $\hat{J}^-$ and $\hat{J}^+$, we obtain the master equation in the form [4]:

$$\dot{\rho} = \frac{\gamma_0}{2}\left(2\hat{J}^- \rho \hat{J}^+ - \hat{J}^+ \hat{J}^- \rho - \rho \hat{J}^+ \hat{J}^-\right) \tag{9}$$

Computer simulation shows (see also Ref. [4]) that the solutions of Eqs. (3) and (9) are similar. They both predict the delay time and the short duration of radiation. Nevertheless, there are substantial differences in maximum values of the intensity and relaxation rates. Moreover, since the right-hand side of Eq. (9) contains the term describing radiation energy loss, an initial pure state should turn into a mixed state so that the Dicke assumption should be violated. The result of computer simulation of Eq. (9) (see Fig. 2) shows that $\mathrm{Tr}\left(\rho^2(t)\right)$, which should be equal to unity for a pure state, deviates from this value indicating that the state becomes mixed and the time evolution goes through a non-Dicke channel.

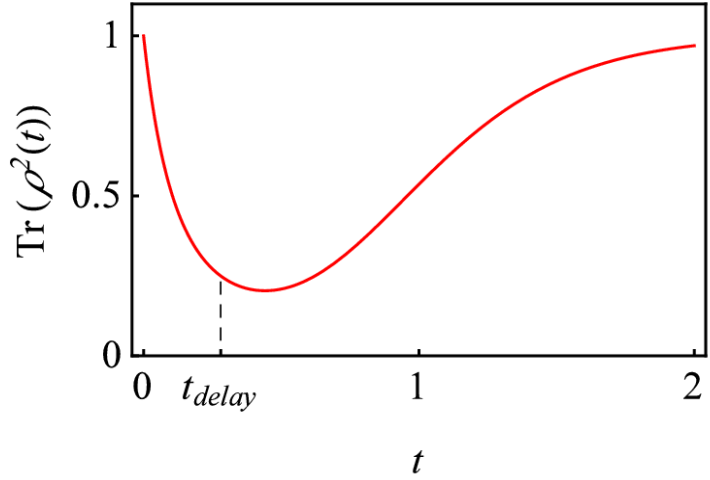

Fig. 2. The dependence of $\text{Tr}(\rho^2(t))$ on time.

### 3. SUPERRADIANCE OF NON-DICKE STATES

In a number of works, the Dicke approach is refined [24-31]. In these papers, it was assumed that dipole moments for both pure and mixed states are zero. In this section, by using Eq. (9) we study a possibility of SR from non-Dicke states with a nonzero dipole moment of each atom.

#### A. Phase operator for a two-level atom

When quantum emitters are excited by pulse pumping, the probability that a two-level atom is in a ground state is nonzero, regardless of the pump power. In general, even within the framework of pure states, the wave function of the final state of a two-level atom is a superposition, $\psi_s = c_e |e\rangle + c_g |g\rangle$, where coefficients $c_e$ and $c_g$ are complex numbers. In this superposition state, an average value of the operator of complex dipole moment $\hat{\sigma}$ is not zero

$$\langle \hat{\sigma} \rangle = \text{Tr}(\hat{\sigma}\rho) = \text{Tr}\left[ |g\rangle\langle e| \left( c_e |e\rangle + c_g |g\rangle \right) \left( c_e^* \langle e| + c_g^* \langle g| \right) \right] = c_e c_g^* \langle g|\hat{\sigma}|e\rangle = \alpha e^{i\varphi}, \quad (10)$$

where $\alpha e^{i\varphi} = c_e c_g^*$ and $\rho = |\psi_s\rangle\langle\psi_s|$. The quantities $\varphi$ and $\alpha$ are analogous to the phase and amplitude of the dipole moment of a classical emitter.

To characterize such a system, we can use the phase of the dipole moment. This is not convenient, though, because we cannot treat in such a manner the Dicke states as their dipole moments are equal to zero. Instead, following Refs. [32-34] in which the phase operator for the photon ensemble has been introduced, we define the phase operator of an *M*-level atom

$$\exp(i\hat{\Phi}) = \left( \sum_{m=0}^{M-2} |m\rangle\langle m+1| \right) + \left( |M-1\rangle\langle 0| \right) \quad (11)$$

For a two-level atom, $M = 2$, we have

$$\cos \hat{\Phi} = \begin{pmatrix} 0 & 1/2 \\ 1/2 & 0 \end{pmatrix}. \quad (12)$$

For an ensemble of $N$ atoms, in the $2^N$-dimensional space, the phase operator $\hat{\Phi}_i$ of an $i$-th atom is defined as a direct product of the operator of the phase of the $i$-th atom, Eq. (12), and unit operators of the other atoms:

$$\cos \hat{\Phi}_i = \hat{I}_1 \otimes ... \otimes \underbrace{\cos \hat{\Phi}}_{i} \otimes ... \otimes \hat{I}_N. \quad (13)$$

The consideration of the time evolution of the expected value of the operator (13) sheds light on the origin of SR. Obviously, an average value of the operator of the phase difference of any two atoms of a system in the Dike state with $n$ excited atoms is equal to zero

$$\langle N, n | \cos \hat{\Phi}_i - \cos \hat{\Phi}_j | N, n \rangle = 0. \quad (14)$$

However, the dispersion of this operator is not zero

$$D_{ij} \equiv \langle N, n | \left( \cos \hat{\Phi}_i - \cos \hat{\Phi}_j \right)^2 | N, n \rangle - \langle N, n | \cos \hat{\Phi}_i - \cos \hat{\Phi}_j | N, n \rangle^2 = \frac{1}{2} - \frac{n(N-n)}{N(N-1)}. \quad (15)$$

As follows from Eq. (15), $D_{ij}$, has a minimum value for $n = N/2$. This is the moment at which SR occurs. Thus, the time of an SR burst can be identified as the moment when the quantum system reaches the phase synchronism, i.e., when the dispersion, $D_{ij}$, is minimal.

### B. Mixed states

Below, to emphasize the difference of our approach from the Dicke model, we consider mixed, non-Dicke states with a nonzero dipole moment as an initial state. The phase operator (11)-(13) can also be applied to such states. To do this, we represent the initial density matrix of non-Dicke states as the direct product of density matrices of individual atoms $\rho = \rho_1 \otimes \rho_2 ... \otimes \rho_N$. The initial density matrix of the $i$-th atom can be represented as,

$$\rho_i = \begin{pmatrix} k_i & \alpha_i e^{i\varphi_i} \\ \alpha_i e^{-i\varphi_i} & 1 - k_i \end{pmatrix}, \quad (16)$$

where $k_i$, $\alpha_i$, and $\varphi_i$ are assumed to be real numbers [35]. Note that the average value of the complex dipole moment $\hat{\sigma}_i$ coincides with the expression obtained for a pure state, Eq. (10):

$$\langle\hat{\sigma}_i\rangle = \text{Tr}(\hat{\sigma}_i\hat{\rho}) = \text{Tr}\left[\begin{pmatrix} 0 & 0 \\ 1 & 0 \end{pmatrix}\begin{pmatrix} k_i & \alpha_i e^{i\varphi_i} \\ \alpha_i e^{-i\varphi_i} & 1-k_i \end{pmatrix}\right] = \alpha_i e^{i\varphi_i}. \quad (17)$$

Thus, the values $\alpha_i$ and $\varphi_i$ have the same physical meanings as in Eq. (10).

The phase operator (11)-(13) can be also applied to non-Dicke states with a nonzero dipole moment. The dynamics of the dispersion of the phase difference is studied by computer simulation of master equation (9). The dimension of the whole system is $2^{2N}$. In our computer simulation, $N \leq 8$, i.e., the order of the system of equations is $2^{16}-1=65,535$. The results are displayed in Fig. 3 where the radiation intensity, defined by Dicke as the time derivative of population inversion, $I(t) = -d\langle \hat{J}^z \rangle / dt$, as well as the dispersion $D_{ij}$ are shown. We can see that in this case, similar to the Dicke case, the SR burst and the minimum of the dispersion happen at the same time.

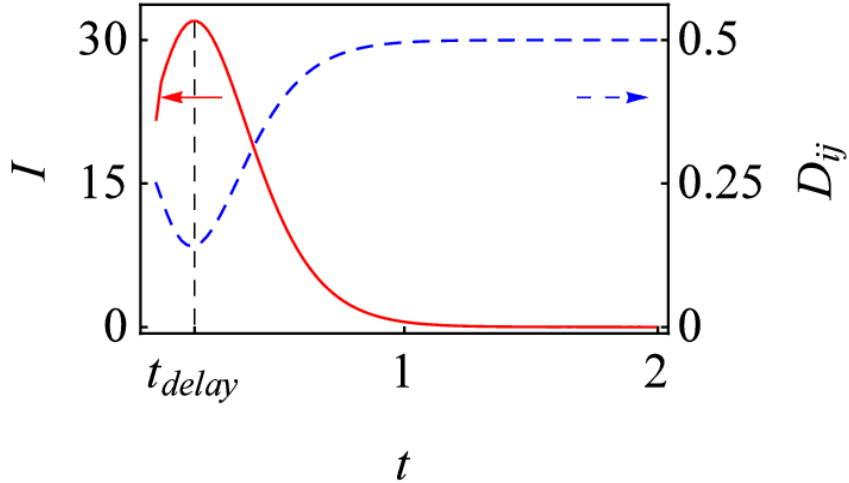

Fig. 3. The dynamics of the intensity $I$ (the solid red line) and the dispersion of the difference of cosines of dipole moment phases (the dashed blue line) for a non-Dicke initial state with a nonzero dipole moment. At the initial moment, eight emitter phases are uniformly distributed in the interval $(-\pi/5, \pi/5)$.

In the case of nonzero dipole moment, it is possible to obtain a relation between the state phase and the phase of a dipole moment. The average value of the cosine of the operator $\hat{\Phi}$ [see Eq. (12)] calculated for mixed state (16) is connected with the phase of the dipole moment $\varphi$ [see Eq. (17)]:

$$\langle\cos\hat{\Phi}\rangle = \text{Tr}\left(\begin{pmatrix} k & \alpha\exp(i\varphi) \\ \alpha\exp(-i\varphi) & 1-k \end{pmatrix}\begin{pmatrix} 0 & 1/2 \\ 1/2 & 0 \end{pmatrix}\right) = \alpha\cos\varphi. \quad (18)$$

Equation (18) shows that that phase of a dipole moment is uniquely related to the state phase. Note that the difference of cosines of phases of dipole moments is equal to average values of the difference of operators of cosines of phases for states of any two atoms are

$$\left\langle \cos \hat{\Phi}_i - \cos \hat{\Phi}_j \right\rangle = \alpha_i \cos \varphi_i - \alpha_j \cos \varphi_j. \tag{19}$$

Equation (19) relates classical and quantum phases.

Thus, if absolute values of dipole moments of atoms are the same, then an average value of the operator of the difference of cosines of phases is proportional to the difference of cosines of dipole moment phases.

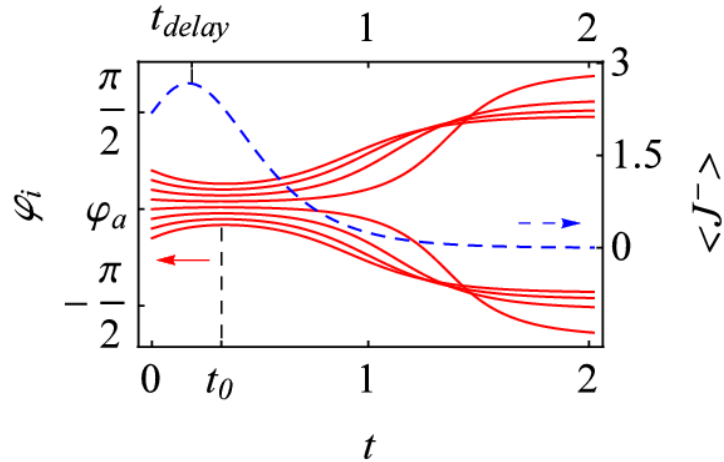

Fig. 4. The dynamics of phases of eight non-Dicke emitters with nonzero dipole moments. The time evolutions of dipole moment phases and the total dipole moment $\left\langle \hat{J}^- \right\rangle$ are shown by solid and dashed lines, respectively.

Let us now consider the dynamics of phases of dipole moments. Our computer simulation shows that at the moment $t_0$, which is near $t_{delay}$, emitter phases become close to each other as shown in Fig. 4. This time coincides with the time at which dispersions of the dipole phase,

$$\Delta = \left( \sum_i \cos \varphi_i^2 - \left( \sum_i \cos \varphi_i \right)^2 / N \right) / (N-1), \tag{20}$$

reach their minimum (see Fig. 5). The duration of the radiation burst is close to the prediction of the Dicke model.

Numerical simulations shows that approaching $t_{delay}$ both dispersions, $D_{ij}$ and $\Delta$, decrease. They reach their minimum near the SR burst, i.e. near the time $t_{delay}$ (see Fig. 5). This is in qualitative agreement with Eq. (19). Similar to classical dipoles [13], the phase convergence shown in Fig. 4 indicates constructive interference in envelopes of the fast dipole oscillations.

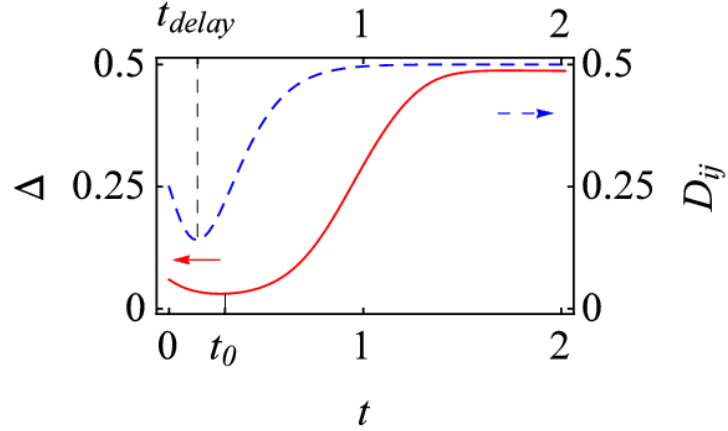

Fig. 5. Solid blue and dashed red lines show dynamics of the dispersions $\Delta$ and $D_{ij}$, respectively. The dispersions are calculated for a non-Dicke initial state with a nonzero dipole moment. At the initial moment, the emitter phases are uniformly distributed in the interval $(-\pi/5, \pi/5)$.

Note that the time $t_{delay}$ is smaller than the time of the phase convergence $t_0$. This happens because the maximum of the total dipole moment is defined by two processes. First, the convergence of phases of emitters of dipole moments leads to constructive interference and to increase in the total dipole moment. Second, during time evolution the system radiates that results in a decrease in the dipole moments. The interplay of these processes leads the SR burst to occur before the phase convergence at time moment $t_{delay}$ where the individual dipole moments are still large enough to form a large total moment.

Thus, we may conclude that SR from non-Dicke states arises due to the phase convergence of emitter dipole moments.

### C. Role of dipole moment

If a classical system of emitters initially oscillates in phase, then similar to SR, the radiation intensity is proportional to the square of the number of particles. However, the delay time of such a system is zero [36-38]. In order to have a delay time, the emitters must have a phase spread [13] which also results in a decrease in the initial dipole moment. This is also true for a quantum system: an increase in the average initial dipole moment leads to a decrease in the delay time. The limiting case of the zero initial dipole moment corresponds to the Dicke model and produces the maximum delay time.

In the Dicke model, for the initial state $|N,n\rangle$ with $N > n > N/2$, as $n$ tends to $N/2$, the delay time approaches zero. Indeed, as follows from Eq. (4), if at the initial moment $\langle \hat{J}_z(0) \rangle = 1$, which corresponds to $n \approx N/2$, then the delay time is zero, $t_{delay} = 0$.

For a non-Dicke state with a nonzero dipole moment, an average number of excited atoms is smaller than $N$. Indeed, an atom can have a nonzero dipole moment if it is in a superposition state, $c_e|e\rangle + c_g|g\rangle$. In this state, an atom has projections on both excited and ground states, and its dipole moment is $|\langle d \rangle| = c_e\sqrt{1-c_e^2}$. Since the probability to find an atom in the excited state is $c_e^2 = \left(1 + \sqrt{1-|\langle d \rangle|^2}\right)/2$, then when the initial dipole moment increases, the initial effective number of excited atoms $N > n_{eff} = Nc_e^2 > N/2$ and, therefore, the delay time should decrease (see Fig. 1). This is confirmed by the results of a numerical experiment shown in Fig. 6.

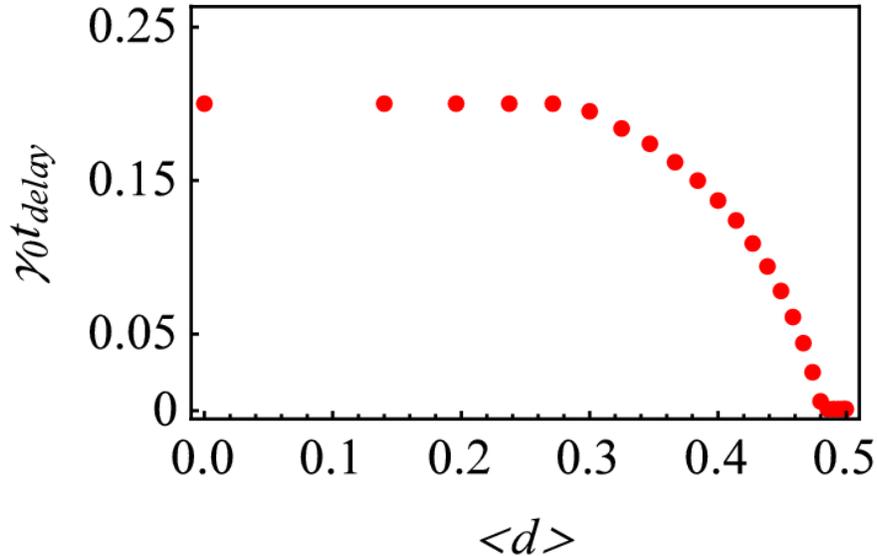

Fig. 6. The dependence of the SR delay time on the average value of the dipole moment per atom.

We can conclude, that there is a close connection between SR in systems of nonlinear classical emitters and two-level atoms. Both systems have attraction points for emitter phases and the delay times decrease when both classical and quantum dipole moments decreases. Since nonlinearity of the classical system is critical for SR, below we show that the same is true for the quantum system.

## D. Role of nonlinearity

Due to the effect of saturation of the population inversion, a two-level emitter is a nonlinear system [39,40]. As we show, this nonlinearity causes the phase conversion of emitters and a decrease in the dispersion of the phase of the emitter state. A many-level system, e.g., a harmonic oscillator, does not superradiate. This is similar to a system of classical oscillators which does not superradiate for a random uniform distribution of phases.

Let us consider whether an SR burst can arise in a system of identical quantum linear harmonic oscillators. We assume that as a system of two-level atoms, oscillators are in a subwavelength volume, and they interact with modes of the electromagnetic field of the free space. This interaction has a form $-\mathbf{d}_i \mathbf{E}$, where $\mathbf{d}_i = \mathbf{d}_0 (\hat{a}_i + \hat{a}_i^+)$. The dynamics of the density matrix is described by the Lindblad equation

$$\dot{\rho} = \frac{\gamma_0}{2} \left( 2\hat{A}\rho\hat{A}^+ - \hat{A}^+\hat{A}\rho - \rho\hat{A}^+\hat{A} \right) \tag{21}$$

where $\hat{A} = \sum_i \hat{a}_i$ is the operator of the collective dipole moment of oscillators, $\hat{a}_i$ is the annihilation operator of the *i*-th oscillator. This equation is derived in a similar way as master equation (9) for a system of two-level atoms.

Equation (21) allows one to obtain dynamics equations for a dipole moment of each oscillator:

$$\langle \dot{\hat{a}}_i \rangle = Tr(\hat{a}_i \dot{\rho}) = \frac{\gamma_0}{2} \langle \hat{a}_i^+ \hat{a}_i \hat{a}_i - \hat{a}_i \hat{a}_i^+ \hat{a}_i \rangle + \frac{\gamma_0}{2} \left\langle \left( \sum_{j \neq i} \hat{a}_j^+ \right) \hat{a}_i \left( \sum_{k \neq i} \hat{a}_k \right) - \hat{a}_i \left( \sum_{j \neq i} \hat{a}_j^+ \right) \left( \sum_{k \neq i} \hat{a}_k \right) \right\rangle$$
$$+ \frac{\gamma_0}{2} \left\langle \left( \sum_{j \neq i} \hat{a}_j^+ \right) \hat{a}_i \hat{a}_i - \hat{a}_i \left( \sum_{j \neq i} \hat{a}_j^+ \right) \hat{a}_i \right\rangle + \frac{\gamma_0}{2} \left\langle \hat{a}_i^+ \hat{a}_i \left( \sum_{k \neq i} \hat{a}_k \right) - \hat{a}_i \hat{a}_i^+ \left( \sum_{k \neq i} \hat{a}_k \right) \right\rangle. \tag{22}$$

Let us estimate each average at the right-hand side of Eq. (22). The first one determines an attenuation of a harmonic oscillator in vacuum:

$$\frac{\gamma_0}{2} \langle \hat{a}_i^+ \hat{a}_i \hat{a}_i - \hat{a}_i \hat{a}_i^+ \hat{a}_i \rangle = -\frac{\gamma_0}{2} \langle \hat{a}_i \rangle, \tag{23}$$

where we use that for a harmonic oscillator the commutator $[\hat{a}_i, \hat{a}_i^+] = \hat{1}$. The second term is zero because the operators corresponding to different oscillators commute. The third and the fourth terms determine attenuation due to collective interaction of oscillators with the modes of free space:

$$\frac{\gamma_0}{2}\left\langle \left(\sum_{j\neq i}\hat{a}_j^+\right)\hat{a}_i\hat{a}_i - \hat{a}_i\left(\sum_{j\neq i}\hat{a}_j^+\right)\hat{a}_i\right\rangle + \frac{\gamma_0}{2}\left\langle \hat{a}_i^+\hat{a}_i\left(\sum_{k\neq i}\hat{a}_k\right) - \hat{a}_i\hat{a}_i^+\left(\sum_{k\neq i}\hat{a}_k\right)\right\rangle$$
$$= -\frac{\gamma_0}{2}\left\langle \left[\hat{a}_i, \hat{a}_i^+\right]\sum_{k\neq i}\hat{a}_k\right\rangle.$$
(24)

Combining Eqs. (23) and (24) and using $\left[\hat{a}_i, \hat{a}_i^+\right] = \hat{1}$ we obtain:

$$\left\langle \dot{\hat{a}}_i\right\rangle = -\frac{\gamma_0}{2}\sum_k \left\langle \hat{a}_k\right\rangle.$$
(25)

Thus, the final system of equation that describes the dynamics of harmonic oscillators is closed with respect to variables $\left\langle \hat{a}_k\right\rangle$. This is a linear system of differential equations for oscillator amplitudes $\left\langle \hat{a}_k\right\rangle$. Since in system (25), all oscillator velocities are the same, any solution of this system attenuates with exactly this velocity. Therefore, there is no SR burst in this system.

## 4. CONCLUSION

We show that there is an analogy between SR in quantum and nonlinear classical systems [13,41,42]. This analogy can be recognized by considering SR from non-Dicke states. In both systems, at the moment of the phase convergence, all dipole moments of the emitters are in phase resulting in an SR burst. The convergence of emitter phases for a system of nonlinear classical emitters arises due to the formation of an attraction point for the phase evolution of the dipole moments of emitters [13]. The existence of an attraction point is a consequence of the nonlinear nature of the process. Our numerical simulations show that an attraction point of phases exists for a system of quantum emitters as well (see Fig. 4). This is likely caused by a nonlinear response of two-level atoms on the electromagnetic field due to the effect of saturation [43].

The behavior of the delay time in a quantum system is also similar to that in a classical system. If the dipole moment of a quantum system initially has its maximum value, then the delay time is zero. In a nonlinear classical system, a system has maximum dipole moment when all emitters are initially in phase. In this case, SR starts without any time delay.

Nonlinearity plays a critical role for SR in both classical and quantum systems. As shown in Ref. [14], in a linear classical system, SR does not occurs. Nonlinearity of a quantum system of two-level atoms is due to their saturation at excitation [39,40]. A system of linear quantum oscillators, which has no saturation, does not superradiate.

To conclude, we have studied the dynamics of quantum emitters interacting via their radiation field. In contrast to the Dicke model, in which all emitters are assumed to be in a state with zero dipole moment, the new SR regime arises in a more realistic system in which the initial state may have a nonzero dipole moment. We demonstrate that the Dicke state is not necessary for SR. Since the Dicke state can be realized only in a limited number of physical systems we expect that our study will stimulate the search for SR which we have shown may be observed in simpler and more realistic systems.

A.A.L would like to acknowledge support from the NSF under Grant No. DMR-1312707.